\documentclass[aps,prd,superscriptaddress,showkeys,nofootinbib,twocolumn,floatfix]{revtex4-2}

\usepackage{graphicx}
\usepackage{multirow}
\usepackage{amsmath,amssymb,amsfonts}
\usepackage{upgreek}
\usepackage{amsthm}
\usepackage{mathtools}
\usepackage[title]{appendix}
\usepackage[table]{xcolor}
\usepackage{textcomp}
\usepackage{booktabs}

\usepackage{hyperref}
\usepackage[capitalize]{cleveref}

\usepackage{todonotes}

\usepackage{tol_colors}

\newcommand\teps{\boldsymbol{\upvarepsilon}}
\newcommand\eps{\varepsilon}
\newcommand\DKL{D_{\textup{KL}}}
\newcommand\DsKL{D_{\textup{sKL}}}
\renewcommand\tensor[1]{\mathbf{#1}}
\newcommand\ttcore[2]{\tensor{#1}^{(#2)}}
\makeatletter
\def\CT@@do@color{%
  \global\let\CT@do@color\relax
        \@tempdima\wd\z@
        \advance\@tempdima\@tempdimb
        \advance\@tempdima\@tempdimc
\advance\@tempdimb\tabcolsep
\advance\@tempdimc\tabcolsep
\advance\@tempdima2\tabcolsep
        \kern-\@tempdimb
        \leaders\vrule
                \hskip\@tempdima\@plus  1fill
        \kern-\@tempdimc
        \hskip-\wd\z@ \@plus -1fill }
\makeatother
\newtheorem{theorem}{Theorem}

\begin{document}

\title[Article Title]{Taming numerical imprecision by adapting the KL divergence to negative probabilities}

% affiliations
\newcommand\URPhysics{Department of Physics, University of Regensburg, 93040 Regensburg, Germany}
\newcommand\URFIDS{Department of Informatics and Data Science, University of Regensburg, 93040 Regensburg, Germany}
\newcommand\ETH{Department of Biosystems and Engineering, ETH Z\"urich,  4056 Basel, Switzerland}
\newcommand\RWTH{Institute for Geometry and Applied Mathematics, RWTH Aachen, 52062 Aachen, Germany}

% author list
\author{Simon Pfahler}\email{simon.pfahler@ur.de}\affiliation{\URPhysics}
\author{Peter Georg}\affiliation{\URPhysics}
\author{Rudolf Schill}\affiliation{\ETH}
\author{Maren Klever}\affiliation{\RWTH}
\author{Lars Grasedyck}\affiliation{\RWTH}
\author{Rainer Spang}\affiliation{\URFIDS}
\author{Tilo Wettig}\email{tilo.wettig@ur.de}\affiliation{\URPhysics}

\date{\today}

\begin{abstract}
    The Kullback-Leibler (KL) divergence is frequently used in data science.
    For discrete distributions on large state spaces, approximations of probability vectors may result in a few small negative entries, rendering the KL divergence undefined.
    We address this problem by introducing a parameterized family of substitute divergence measures, the shifted KL (sKL) divergence measures.
    Our approach is generic and does not increase the computational overhead.
    We show that the sKL divergence shares important theoretical properties with the KL divergence and discuss how its shift parameters should be chosen. 
    If Gaussian noise is added to a probability vector, we prove that the average sKL divergence converges to the KL divergence for small enough noise. 
    We also show that our method solves the problem of negative entries in an application from computational oncology, the optimization of Mutual Hazard Networks for cancer progression using tensor-train approximations.
\end{abstract}


\maketitle

\section{Introduction}
\subsection{Motivation}
A notion of distance between probability distributions is required in many fields, e.g., information and probability theory, approximate Bayesian computation~\cite{Csillery_2010,Schill_2019}, spectral (re-) construction~\cite{Hoch_2014,Rothkopf_2017,Ha_2019}, or compression using variational autoencoders~\cite{Kingma_2022}.
In mathematical terms, this notion can be phrased in terms of divergence measures.

One of the most commonly used divergence measures is the Kullback-Leibler (KL) divergence~\cite{Kullback_1951}, which for two discrete probability vectors $\tensor p=(p_1,\ldots,p_n)$ and $\tensor q=(q_1,\ldots,q_n)$ has the form
\begin{equation}\label{eq:KL}
    \DKL(\tensor{p}\|\tensor{q})=\sum_{i=1}^n p_i\log\frac{p_i}{q_i}\,.
\end{equation}
The KL divergence includes the logarithm, which is well-defined in the usual case of probability vectors with nonnegative entries.
However, in practical applications, situations can arise that lead to negative entries, for which the logarithm is not defined.
Let us give a few examples.
In high-dimensional spaces, approximations are often needed to keep calculations tractable and to avoid runtime explosion due to the curse of dimensionality~\cite{Thomas_2015,Georg_2022}.
Such approximations can lead to negative entries when the exact entry is close to zero, as is frequently the case for probability vectors.
Another source of negative entries are rounding errors that occur, e.g., when the probability vector is obtained as the solution of a linear system of equations~\cite{Philippe_1992,Schill_2019}.
Also, assumptions of a theory can lead to negative entries in the approximate probability distribution~\cite{Alkofer_2001,Burnier_2011,Rothkopf_2017}.

If negative entries arise, one is faced with the choice of (a) giving up, (b) reformulating the theory, model, or approximation such that negative entries cannot occur, or (c) devising a workaround that leads to correct results in a suitable limit.
In ~\cref{sec:related} we review approaches from category (b) and (c) to address the problem of negative entries~\cite{Paatero_1994,Hobson_1998,Welling_2001,Chi_2012,Haas_2014,Hansen_2015,Lee_2016,Rothkopf_2017}.
The methods we are aware of are typically tailored to the problem at hand or lead to reduced run-time performance.

Here, we suggest a method in category (c) that is generically applicable and does not suffer from decreased performance.
Specifically, we propose the \emph{shifted Kullback-Leibler (sKL) divergence} as a substitute divergence measure for the standard KL divergence in the case of approximate probability vectors with negative entries.
It is parameterized by a vector of shift parameters.
The sKL divergence is a modification of the KL divergence and retains many of the latter's useful properties while handling negative entries when they arise.
To give theoretical support to our method, we consider a simple example in which i.i.d.\ Gaussian noise is added to the entries $q_i$ in \cref{eq:KL} so that negative entries can occur.
For this example we prove that the difference between the KL divergence and the expected sKL divergence under the noise is quadratic in the standard deviation of the noise, provided that the shift parameters are suitably chosen.
Therefore the sKL divergence converges to the KL divergence in the limit of small i.i.d.\ Gaussian noise.
We also show this convergence numerically.

In a concrete application, we show that the sKL divergence enables efficient learning of a Mutual Hazard Network (MHN), a model of cancer progression as an accumulation of certain genetic events~\cite{Schill_2019,Chen_2023,Luo_2023}.
In an MHN, cancer progression is modeled as a continuous-time Markov chain whose transition rates are parameters of the model that are learned such that the model explains a given patient-data distribution.
When considering $\gtrsim 25$ possible genetic events, exact model construction becomes impractical and requires efficient approximation techniques~\cite{Georg_2022}.
We show that even though negative entries occur in the approximate probability vectors, meaningful models can still be constructed when the sKL divergence is employed.

\subsection{Related work} \label{sec:related}
Various approaches to handling or avoiding negative entries in approximate probability distributions exist in the literature.

In earlier work~\cite{Georg_2022_PhD}, we considered introducing a threshold $\eps > 0$ and replacing entries of the probability vector that are smaller than this threshold by $\eps$.
This method fails to preserve probability mass and leads to a loss of gradient information when used in an optimization task.
Additionally, the comparison function obtained in this way no longer meets the requirements of a statistical divergence~\cite{Amari_2016}, i.e., none of the required properties in \cref{thm:divergence} below are satisfied in this case.

For high-dimensional probability distributions, several nonnegative tensor decompositions have been proposed in order to break the curse of dimensionality while avoiding negative entries~\cite{Paatero_1994,Welling_2001,Chi_2012,Hansen_2015,Lee_2016}.
Typically, this change of format leads to a significant increase in the run time of the algorithms involved, as other important properties for a time-efficient approximation have to be omitted.

In situations where negative entries occur due to assumptions of the theory, a common approach is to modify the divergence measure~\cite{Hobson_1998,Rothkopf_2017,Haas_2014}.
Such modifications are typically problem-specific and therefore not applicable in general.

Our approach is in a similar spirit, but our modification of an established divergence measure is not problem-specific, as it does not depend on special properties of a particular application.
Furthermore, since we do not need specific formats to preserve nonnegativity, optimization tasks in high dimensions can be completed efficiently.

This paper is structured as follows. In \cref{sec:theory} we define the sKL divergence, discuss its properties and give suggestions for how to choose the shift parameters.
In \cref{sec:application} we show how the sKL divergence can be applied in optimization tasks on approximate probability vectors.
In \cref{sec:outlook} we summarize our findings and give an outlook on future research topics.
In \cref{sec:proofs,sec:Q_matrix_TT} we provide proofs of three theorems on the sKL divergence and mathematical details of the approximation we employ.

\section{Theory} \label{sec:theory}
\begin{figure}[t]
    \centering
    \includegraphics{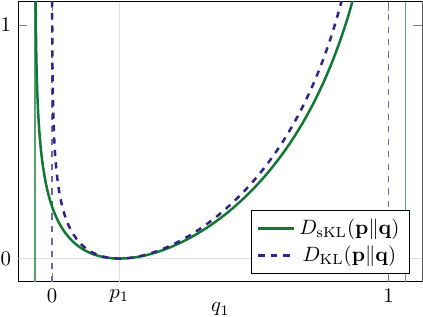}
    \caption{Plot of the KL and sKL divergence of probability vector $\tensor{p}=\left(0.2, 0.8\right)$ from $\tensor{q}=\left(q_1, 1-q_1\right)$. For the sKL divergence, $\eps_i=0.05$ was chosen for $i=1,2$.}
    \label{fig:1D_sKL_plot}
\end{figure}

\subsection{The sKL divergence}
For two probability vectors $\tensor{p}$ and $\tensor{q}$ on a finite set of $n$ elements, the Kullback-Leibler (KL) divergence of $\tensor{p}$ from $\tensor{q}$ is defined as~\cite{Kullback_1951}
\begin{equation*}
    \DKL(\tensor{p}\|\tensor{q})=\sum_{i=1}^n p_i\log\frac{p_i}{q_i}
\end{equation*}
with the convention $0 \log(0/x)=0$.
In the context of statistical modeling, the vector $\tensor{p}$ (with $p_i\geq0$) is typically given by the data, while the vector $\tensor{q}$ is a theoretical model.
This model can be fitted to the data by minimizing the KL divergence of $\tensor{p}$ from $\tensor{q}$.
Approximations in the calculation of $\tensor{q}$ can lead to negative entries $q_i<0$.%
\footnote{
    For simplicity, we do not consider the possibility that the entries $q_i$ are zero up to machine precision because this is extremely unlikely to happen in a numerical simulation. However, it is straightforward to extend our analysis to this case.
}
In this case, the KL divergence is no longer defined and the optimization cannot be done.
For this scenario, we propose the shifted Kullback-Leibler (sKL) divergence, a modification of the Kullback-Leibler divergence that allows for negative entries in $\tensor{q}$.
For a parameter vector $\teps\in\mathbb{R}_{\geq 0}^n$, we define it as
\begin{equation}
    \DsKL(\tensor{p}\|\tensor{q})=\sum_i \left(p_i+\eps_i\right)\log\frac{p_i+\eps_i}{q_i+\eps_i}
\end{equation}
with the same convention as above.
The sKL divergence is well defined for $q_i>-\eps_i$.
It is a modification of the KL divergence that reduces to the KL divergence for $\teps=0$.
\Cref{fig:1D_sKL_plot} shows the behavior of both the KL and the sKL divergence for probability vectors on a set with two possible outcomes.
We can see that the domain on which the sKL divergence is defined now also includes approximate probability vectors with small negative entries.
In the following two subsections we show that the sKL divergence retains many important properties of the KL divergence and discuss how to best choose the parameters $\eps_i$.

\subsection{Properties} \label{sec:theory:properties}
The KL divergence belongs to the class of $f$-divergences~\cite{Basseville_2013}.
While the sKL divergence does not, it shares many important properties with the KL divergence: it is positive semidefinite, only zero for $\tensor{p}=\tensor{q}$, and locally a metric. 
Thus, it still satisfies the definition of a statistical divergence, see \cref{thm:divergence}.

\smallskip

\begin{theorem} \label{thm:divergence}
    Let $\tensor{p}$, $\tensor{q}$ and $\teps$ be three vectors in $\mathbb{R}^n$ whose entries satisfy $p_i,q_i>-\eps_i$ and $\sum_ip_i=\sum_iq_i$.
    The sKL divergence of $\tensor{p}$ from $\tensor{q}$, with parameter vector $\teps$, is a statistical divergence, i.e., it satisfies the following properties~\cite{Amari_2016},
    \begin{enumerate}
        \item[\hypertarget{thm:div1}{(1)}] $\DsKL(\tensor{p}\|\tensor{q})\geq 0\,,$
        \item[\hypertarget{thm:div2}{(2)}] $\DsKL(\tensor{p}\|\tensor{q})=0$ if and only if $\tensor{p}=\tensor{q}\,,$
        \item[\hypertarget{thm:div3}{(3)}] $\frac{\textup{d}^2\DsKL(\tensor{p}\|\tensor{q})}{\textup{d}q_i\textup{d}q_j}\big|_{\tensor{q}=\tensor{p}}$ is a positive definite matrix.
    \end{enumerate}
\end{theorem}
\smallskip
\noindent
The first assumption of the theorem can always be satisfied by a suitable choice of the shift parameters $\eps_i$, while the second assumption can always be satisfied by a rescaling of the $q_i$. 
Additionally, we show in \cref{thm:convexity} that the sKL divergence is convex in the pair of its arguments, like the KL divergence~\cite{vanErven_2014}.

\smallskip

\begin{theorem} \label{thm:convexity}
    For fixed parameter vector $\teps$, the sKL divergence is convex in the pair of its arguments.
    That is, if $(\tensor{p}^{(1)}, \tensor{q}^{(1)})$ and $(\tensor{p}^{(2)}, \tensor{q}^{(2)})$ are two pairs of vectors in $\mathbb{R}^n$ for which $\DsKL(\tensor{p}^{(1)}\|\tensor{q^{(1)}})$ and $\DsKL(\tensor{p}^{(2)}\|\tensor{q^{(2)}})$ are well-defined, and if $\lambda\in[0, 1]$, it satisfies
    \begin{align*}
        &\DsKL(\lambda \tensor{p}^{(1)}+(1-\lambda)\tensor{p}^{(2)}\|\lambda \tensor{q}^{(1)}+(1-\lambda)\tensor{q}^{(2)}) \\
        \quad &\leq \lambda \DsKL(\tensor{p}^{(1)}\|\tensor{q}^{(1)}) + (1-\lambda) \DsKL(\tensor{p}^{(2)}\|\tensor{q}^{(2)})\,.
    \end{align*}
\end{theorem}
\smallskip
\noindent 
This property is of particular importance as it is often needed for well-behaved optimization~\cite{Fletcher_2000}.
The proofs of the two theorems are given in \cref{sec:proofs}.

\subsection{Parameter choice} \label{sec:theory:parameter_choice}
In the sKL divergence, we introduced a parameter vector $\teps=(\eps_1,\ldots,\eps_n)$.
The properties of the sKL divergence discussed in the previous subsection (\cref{thm:divergence,thm:convexity}) hold for any choice of the shift parameters $\eps_i$.
However, in applications, their values can have a large influence on the quality of the results, such as speed of convergence or accuracy of the learned model.
In this section we therefore aim to provide a guide for choosing suitable values of the parameters $\eps_i$ for the case that $\tensor{p}$ is an exact probability vector and $\tensor{q}$ is an approximate one.
In the KL divergence, the terms in the sum only need to be evaluated for the indices $i$ for which $p_i\neq 0$.
This property reduces the computational cost greatly if many entries of $\tensor{p}$ are zero.
This situation is encountered, for example, when optimizing an MHN~\cite{Schill_2019}.
In order to preserve this advantage when working with the sKL divergence, one can simply choose $\eps_i=0$ if $p_i=0$.
To constrain possible values for $\eps_i$, we first note that if $q_i<0$, the sKL divergence is only well-defined if one chooses $\eps_i>-q_i$.
Second, we compute the gradient of the sKL divergence,
\begin{equation*}
    \frac{\text{d}\DsKL(\tensor{p}\|\tensor{q})}{\text{d}q_i}=-\frac{p_i+\eps_i}{q_i+\eps_i}\,,
\end{equation*}
and notice that it approaches $-1$ for large values of $\eps_i$.
Most optimization algorithms use gradient information to minimize a loss function~\cite{Fletcher_2000}.
Therefore it is important to retain the gradient information, which implies that $\eps_i$ should not be too large.

Using these guidelines, we provide two particular choices of $\teps$.
First, in \cref{sec:theory:static}, we discuss a natural, static choice of $\teps$ and examine its shortcomings.
In \cref{sec:theory:dynamic}, we then suggest a dynamic choice of $\teps$ and prove in \cref{thm:noise} that the resulting sKL divergence is a good substitute for the KL divergence in the presence of small i.i.d.\ Gaussian noise.
When possible, one should therefore prefer the dynamic choice of $\teps$, as we will also demonstrate in \cref{sec:application:results}.

\subsubsection{Static choice} \label{sec:theory:static}
We first suggest a choice of $\teps$ that can be used with any optimization algorithm.
This includes higher-order algorithms such as BFGS or conjugate-gradient methods, which make use of approximations of the Hessian matrix of the loss function~\cite{Fletcher_2000}.
For higher-order methods, the objective function must not change during optimization.
This requires us to choose a suitable $\teps$ a priori, which can be a difficult task.
A simple approach is to evaluate $\tensor{q}$ once before starting the optimization and to set
\begin{equation}\label{eq:static_eps}
    \eps_i=\begin{cases}
        0\,, & p_i=0\,, \\
        \eps\,, & \text{else}\,,
    \end{cases}
\end{equation}
where $\eps$ is fixed to a number slightly larger than $\max_{p_j > 0} (-q_j)$.
If during optimization $\eps$ turns out to be too small, i.e., negative entries of $q_i+\eps$ are encountered, the optimization has to be stopped early.
The best possible result can then be found by tuning $\eps$.

This parameter choice allows us to choose a wide variety of optimizers. 
However, in addition to the requirement of tuning $\eps$, the static choice leads to an unnecessarily large loss of gradient information.
This is because gradient information is reduced for all $i$, even though $q_i<0$ is rarely encountered in practice.

\subsubsection{Dynamic choice} \label{sec:theory:dynamic}
We now suggest a second choice that can be used only if the objective function is allowed to change in every iteration of the optimizer.
This does not pose a problem when using first-order optimization algorithms such as gradient descent~\cite{Fletcher_2000} or Adam~\cite{Kingma_2017}, as these do not use information about the Hessian matrix of the loss function.
In this scenario, we can choose a new $\teps$ for every new $\tensor{q}$ during the optimization process.
Our proposed choice is
\begin{equation}\label{eq:dynamic_eps}
    \eps_i=\begin{cases}
        0\,, & p_i=0 \text{ or } q_i>0\,, \\
        |q_i| + f(|q_i|)\,, & \text{else}\,,
    \end{cases}
\end{equation}
where $f:\mathbb{R}_{\geq0} \rightarrow \mathbb{R}_{\geq0}$ is a nonnegative function.
This choice of $\teps$ avoids a large loss in gradient information while extending the domain on which the sKL divergence is defined where necessary.
A concrete choice of $f$, which we use in \cref{sec:application}, is given by $f(x)=\delta \cdot x$ with a constant $\delta>0$, but many other choices are possible.

\subsection{Negative entries due to Gaussian noise}
To further motivate \cref{eq:dynamic_eps}, let us consider a simple example in which negative entries of $q_i$ are caused by small Gaussian noise.
In this case, we show that the difference between the KL divergence and the average of the sKL divergence over the Gaussian noise is quadratic in the standard deviation of the noise, see \cref{thm:noise}, the proof of which is provided in \cref{sec:proofs}.
To keep the presentation simple, we restrict ourselves to nonnegative functions $f$ that are finite sums of power laws, i.e., $f(x)=\sum_ka_kx^{b_k}$ with at least one $a_k\ne 0$.

\medskip

\begin{theorem} \label{thm:noise}
    Let $\tensor{p}$ and $\tensor{q}$ be two probability vectors on a finite set of $n$ elements, with $q_i>0$ whenever $p_i>0$.
    Let further $\tensor{x}$ be a vector of i.i.d.\ Gaussian random variables with mean $0$ and standard deviation $\sigma$.
    If the $\eps_i$ are chosen according to \cref{eq:dynamic_eps} with the restriction on $f$ stated above, the average of the sKL divergence of $\tensor{p}$ from $\tensor{q}+\tensor{x}$ is given by
    \begin{align*}
        & \big\langle\DsKL(\tensor{p}\|\tensor{q}+\tensor{x})\big\rangle_\tensor{x} \\
        & \quad= \DKL(\tensor{p}\|\tensor{q})+\sigma^2\sum_i\frac{p_i}{2q_i^2}+\mathcal{O}(\sigma^4)\,.
    \end{align*}
\end{theorem}

\medskip
\noindent
In fact, the restriction on $f$ can be relaxed significantly so that a much larger class of functions is admissible, see \cref{proof:noise} for details.

In this example, we therefore see explicitly that the sKL divergence can be used as a substitute divergence measure for the KL divergence, provided that the condition $\sigma^2\sum_ip_i/2q_i^2\ll \DKL(\tensor{p}\|\tensor{q})$ is satisfied.

\begin{figure}[t]
    \centering
    \includegraphics{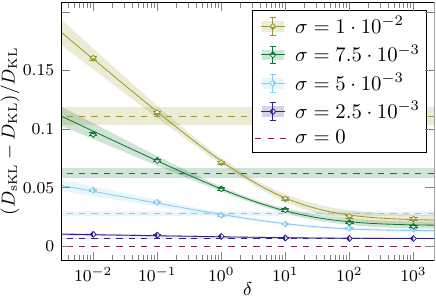}
    \caption{Relative difference between the KL divergence and the sKL divergence with the dynamic choice of $\teps$ given by \cref{eq:dynamic_eps} with $f(x)=\delta\cdot x$, for different noise levels $\sigma$.
    For each $\sigma$, the dashed line shows the prediction from \cref{thm:noise} truncated at order $\sigma^2$, the solid line shows the result obtained from numerical integration of \cref{eq:expected_sKL}, and the points show the result obtained from simulation.}
    \label{fig:sKLconv_Gaussian}
\end{figure}

To validate \cref{thm:noise} numerically, we first consider a generic toy model where $\tensor p,\tensor q\in\mathbb R^{10}$ are probability vectors whose entries are i.i.d.\ uniform random variables.
As per the assumptions of \cref{thm:noise}, we consider i.i.d.\ Gaussian random variables added to $\tensor q$, and we choose the $\eps_i$ according to \cref{eq:dynamic_eps} with the simple choice $f(x)=\delta\cdot x$.
In order for the $\sigma^2$ correction to be small, we restrict ourselves to probability vectors $\tensor p$ and $\tensor q$ that satisfy
\begin{equation*}
    \sigma^2\sum_i\frac{p_i}{2q_i^2}\le \frac12\DKL(\tensor p\|\tensor q)
\end{equation*}
for all $\sigma$ considered.
\Cref{fig:sKLconv_Gaussian} shows the convergence of the sKL divergence to the KL divergence in the limit of small $\sigma$ of the noise.
The plot shows averages of $200$ pairs of $\tensor p$ and $\tensor q$.
We can see that in this case \cref{thm:noise} gives a good prediction for the relative difference between the sKL divergence and the KL divergence.
Interestingly, for large values of $\delta$, the relative difference is significantly lower than predicted.
Therefore, large values $\delta\ge10^1$ should be preferred in this toy model, but this may be different for other applications, in particular when the negative entries are not due to Gaussian noise.
Finally, we observe that the parameter $\delta$ in the function $f$ has only little effect on the result when $\sigma$ is sufficiently small.

In many applications, the strength of the noise can be controlled by parameters that determine the numerical accuracy of the approximation.
If the assumption of i.i.d.\ Gaussian noise is satisfied in a particular application, \cref{thm:noise} is directly applicable.
In other applications, the noise might follow a different distribution, for which one would have to do a similar analysis.

In the concrete application we consider in \cref{sec:application}, we have seen numerically that the assumption of i.i.d.\ Gaussian noise is not satisfied.
Nevertheless, as we increase the numerical accuracy of our approximation, we observe that the sKL divergence with the choice of \cref{eq:dynamic_eps} converges to the KL divergence computed without approximations.
Also, the model results obtained from optimizing the sKL divergence with the choice of \cref{eq:dynamic_eps} converge to the model result obtained from the KL divergence without approximations.

\section{Application} \label{sec:application}
\subsection{Mutual Hazard Networks}
As a real-world application, we consider the modeling of cancer progression using Mutual Hazard Networks (MHNs)~\cite{Schill_2019}.
Cancer progresses by accumulating genetic events such as mutations or copy-number aberrations~\cite{Michor_2004}.
As each event can be either absent or present, the state of an event is represented by $0$ (absent) or $1$ (present).
We consider $d$ such events, thus representing a tumor as a $d$-dimensional vector $x\in\{0,1\}^d$.
An MHN aims to infer promoting and inhibiting influences between single events.
The data used for the learning process are patient data of observed tumors.
The data distribution $\tensor{p}$ is thus a discrete probability distribution on the $2^d$-dimensional state space $\mathcal{S}=\left\{x\in\{0,1\}^d\right\}$ of possible tumors, i.e., $n=2^d$ in the notation of \cref{sec:theory}.

\begin{figure}[t]
    \centering
    \includegraphics{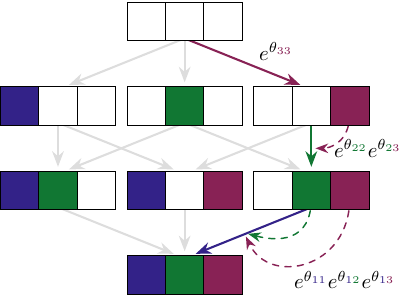}
    \caption{Allowed transitions, along with transition rates, for an MHN with three possible events and parameter matrix $\theta \in \mathbb{R}^{3 \times 3}$.}
    \label{fig:MHN_transitions}
\end{figure}

An MHN models the progression as a continuous-time Markov chain under the assumptions that at time $t=0$ no tumor has active events, that events occur only one at a time, that events are not reversible, and that transition rates follow the Proportional Hazard Assumption~\cite{Cox_1972}.
If we have two states $x, x^{+i}\in\mathcal{S}$ that differ only by $x_i=0$ and $x^{+i}_i=1$, the transition rate from state $x$ to state $x^{+i}$ is modeled as%
\footnote{
    All other off-diagonal transition rates are zero by assumption of the MHN.
}
\begin{equation*}
    \tensor{Q}_{x^{+i}, x} = e^{\theta_{ii}}\prod_{x_j=1}e^{\theta_{ij}}\,,
\end{equation*}
where $e^{\theta_{ii}}$ is the base rate of event $i$ and $e^{\theta_{ij}}$ is the multiplicative influence event $j$ has on event $i$.
An MHN with $d$ events can thus be described by a parameter matrix $\theta\in\mathbb{R}^{d\times d}$.
\Cref{fig:MHN_transitions} shows all allowed transitions and their rates for $d=3$.
The transition-rate matrix can efficiently be written as a sum of $d$ Kronecker products,
\begin{equation}\label{eq:MHN_transition_rate_matrix_kron}
    \tensor{Q}=\sum_{i=1}^d\bigotimes_{j=1}^dQ_{ij}
\end{equation}
with
\begin{equation*}
    Q_{ij}=\begin{pmatrix}1&0\\0&e^{\theta_{ij}}\end{pmatrix}\text{ for } i\neq j    \text{ and }
    Q_{ii}=\begin{pmatrix}-e^{\theta_{ii}}&0\\ \phantom{-}e^{\theta_{ii}}&0\end{pmatrix}.
\end{equation*}

\noindent
Starting from the initial distribution $\tensor{q}_\varnothing=(1,0,0,\ldots)\in\mathbb{R}^{2^d}$, the probability distribution at time $t \geq 0$ is given by Markov-chain theory as
\begin{equation*}
    \tensor{q}(t)=e^{t\tensor{Q}}\tensor{q}_\varnothing\,,
\end{equation*}
where $e^{t\tensor{Q}}$ denotes the matrix exponential.
Since tumor age is not known in the data, MHNs assume that the age of tumors in $\tensor{p}$ is an exponentially distributed random variable with mean $1$.
If we marginalize $\tensor{q}(t)$ over $t$ accordingly, we obtain
\begin{equation}\label{eq:time_marginal_distribution}
    \tensor{q}_\theta \coloneqq \int_0^\infty \text{d}t\;e^{-t}e^{t\tensor{Q}}\tensor{q}_\varnothing=\left(\tensor{I}-\tensor{Q}\right)^{-1}\tensor{q}_\varnothing\,,
\end{equation}
where $\tensor{I}$ denotes the identity.

A parameter matrix $\theta$ that best fits the data distribution $\tensor{p}$ can now be obtained by minimizing a divergence measure of $\tensor{p}$ from $\tensor{q}_\theta$.
For small $d$ (i.e., $d\lesssim25$), the KL divergence can be used since the calculation of $\tensor{q}_\theta$ can be done without approximation.
For larger $d$, this is no longer possible because of the exponential complexity (recall that the dimension of the state space is $2^d$), and the approach has to be modified~\cite{Georg_2022}.
One possible modification is the use of low-rank tensor methods~\cite{Grasedyck_2013} in order to keep calculations tractable.
In particular, we use the tensor-train (TT) format~\cite{Oseledets_2011}, which usually minimizes the approximation error through the Euclidean distance.
Specifically, when approximating a tensor $\tensor{a}$ by $\tilde{\tensor{a}}$, a maximum Euclidean distance $\Delta=||\tensor{a}-\tilde{\tensor{a}}||_2$ can be specified.
Thus, small negative entries can occur in $\tilde{\tensor{a}}$ if the corresponding entry in $\tensor{a}$ is smaller than $\Delta$.
In this case, the KL divergence is no longer defined.
To be able to perform the optimization, we switch to the sKL divergence as our objective function.
In \cref{sec:application:tt}, we explain the basics of the TT format.
\Cref{sec:application:results} shows how we can find suitable $\theta$ matrices by use of the sKL divergence even when negative entries are encountered.

\subsection{Tensor Trains} \label{sec:application:tt}
For a large number $d$ of events, storing $\tensor{q}$ (and $\tensor{p}$) as a $2^d$-dimensional vector is computationally infeasible.
This storage requirement can be alleviated through use of the TT format~\cite{Oseledets_2011}.
A $d$-dimensional tensor $\tensor{a}\in\mathbb{R}^{n_1\times\cdots\times n_d}$ with mode sizes $n_k\in\mathbb{N}$ can be approximated in the TT format as
\begin{equation*}
    \tensor{a}(i_1, \hdots, i_d) \approx
        \sum_{\alpha_0,\hdots,\alpha_d}
        \prod_{k=1}^{d}
        \ttcore{a}{k}(\alpha_{k-1}, i_k, \alpha_k)
\end{equation*}
for all $i_1, \dots, i_d$ with tensor-train cores $\ttcore{a}{k}\in\mathbb{R}^{r_{k-1}\times n_k\times r_k}$, where the $r_k$ are tensor-train ranks.%
\footnote{
    Strictly speaking, the $r_k$ are only called TT ranks if they are chosen minimally.
    To simplify the presentation we use the term ``TT rank'' regardless of this convention.
}
In order to represent a scalar by the right-hand side, the condition $r_0=r_d=1$ is required.
The quality of approximation can be controlled through the TT ranks $r_k$.
In particular, it can be shown that choosing the TT ranks large enough gives an exact representation of $\tensor{a}$~\cite{Oseledets_2011}.

The TT format not only allows for efficient storage of high-dimensional tensors, but also supports many basic operations, e.g., addition, inner products, or operator-by-tensor products~\cite{Oseledets_2011}.
Furthermore, there are efficient algorithms for solving linear equations~\cite{Holtz_2012,Dolgov_2014}.

By modifying the shape of the objects in \cref{eq:MHN_transition_rate_matrix_kron} and changing the Kronecker products to tensor products, the transition-rate matrix $\tensor{Q}\in \mathbb{R}^{\mathcal{S}\times\mathcal{S}}$ can be written in the tensor-train format~\cite{Hackbusch_2019} (for details see \cref{sec:Q_matrix_TT}).
This leads to a tensor train with all TT ranks $r_1,\hdots,r_{d-1}$ equal to $d$.
Thus, we can approximately solve \cref{eq:time_marginal_distribution} in the TT format using~\cite{Holtz_2012}.
Similar techniques can be used for the gradient calculation~\cite{Georg_2022_PhD}.

Details of the algorithms involved~\cite{Holtz_2012} lead to the conditions $r_{k+1}\leq n_{k+1}r_k$ and $r_{k-1}\leq n_kr_k$ on the TT ranks of $\tensor{q}_\theta$.
As a result, the first and last TT ranks are $r_1\le n_1$ and $r_{d-1}\le n_d$.
Towards the middle of the tensor train, ranks increase until they level off at the specified maximum rank.
In our simulations, we specify a maximum TT rank $r_{\tensor{q}}$ and choose the TT ranks $r_k\le r_{\tensor{q}}$ to be as large as possible given the constraints on the $r_k$.

\subsection{Simulations} \label{sec:application:results}
We test how well an MHN can learn a probability distribution $\tensor{p}$ when optimizing the sKL divergence instead of the classical KL divergence.
We use simulated data for $d=20$ events.
This relatively small value of $d$ allows for exact calculations so that we can compare results obtained with and without the use of tensor trains.
In the following, we describe how the data were generated, how the MHNs were learned, and how their quality was assessed.

Every dataset was generated from a ground-truth model described by $\theta_{\text{GT}}\in\mathbb{R}^{d\times d}$.
The diagonal entries of $\theta$ were drawn from a Gaussian distribution $\exp(-(x-\mu)^2/2\sigma^2)/\sigma\sqrt{2\pi}$ with $\mu=-1$ and $\sigma=1$.
A random set of $10\%$ of the off-diagonal entries of $\theta$ was drawn from a Laplace distribution $\exp(-|x-\mu|/b)/2b$ with $\mu=0$ and $b=1$, and the remaining entries were set to $0$.
These choices were made to mimic MHNs obtained from real data we studied.
The data distribution $\tensor{p}$ was obtained from $\theta_{\text{GT}}$ by drawing $1000$ samples from its time-marginalized probability distribution, as defined in \cref{eq:time_marginal_distribution}.

Given a dataset, MHNs were learned by optimizing the sKL divergence.
We indirectly control the magnitude of the largest negative entries of $\tensor{q}_\theta$ through our choice of the maximum possible TT rank $r_\tensor{q}$ of $\tensor{q}_\theta$.
$10$ datasets were generated, and for all of them, MHNs were learned for specific choices of $r_\tensor{q}$ and $\teps$.
The results shown below are arithmetic means from these $10$ runs.
To avoid overfitting, an L1 penalty term of the form $\lambda\sum_{i\neq j}|\theta_{ij}|$ was added to the objective function.
The factor $\lambda$ was not made part of the optimization but set to a constant value of $10^{-3}$ for all simulation runs to make comparison of different models easier.

A learned MHN's quality was assessed using the KL divergence of the ground truth model's time-marginalized probability distribution from the time-marginalized probability distribution of the learned MHN, see \cref{eq:time_marginal_distribution}.
This KL divergence was calculated without the use of the TT format to ensure that no negative entries can occur.

\subsubsection{Static choice} \label{sec:application:static_choice}
\begin{figure}[t]
    \centering
    \includegraphics{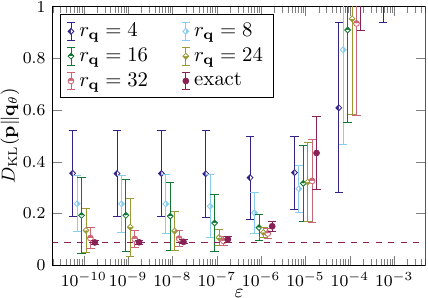}
    \caption{Visualization of the data in \cref{table:MHNstatic}.
        The dashed line indicates the result when optimizing the KL divergence without the use of the TT format.}
    \label{fig:MHNstatic}
\end{figure}

First, we consider the static choice of $\teps$ given in \cref{eq:static_eps}.
In this case, higher-order optimizers can be used for faster convergence to an optimum.
However, for a fair comparison with the dynamic choice, we used gradient descent (a first-order optimizer) for all optimizations.
\Cref{table:MHNstatic,fig:MHNstatic} show the results for various combinations of $r_\tensor{q}$ and $\eps$.
In the last column (``exact''), optimization using the sKL divergence was also done when the TT format was not used, even though the KL divergence is well-defined and the introduction of a positive $\eps$ is not necessary in this case.
The additional entries are written in grey to indicate this.

\begin{table}[t]
    \newcommand\col[1][]{\cellcolor{Trose!#10}}%
\centering
\tabcolsep=5.5pt
\begin{tabular}{c ccccc c}
    \toprule
     & \multicolumn{5}{c}{$r_\tensor{q}$} &  \\\cmidrule{2-6}
    $\eps$ & 4 & 8 & 16 & 24 & 32 & exact \\\midrule
    $0$ & \col[10]0.354 & \col[10]0.241 & \col[10]0.199 & \col[6]0.131 & \col[5]0.106 & 0.087 \\\midrule
    $10^{-10}$ & \col[10]0.355 & \col[10]0.237 & \col[9]0.192 & \col[5]0.135 & \col[3]0.105 & {\color{gray}0.088} \\
    $10^{-9}$ & \col[10]0.354 & \col[10]0.237 & \col[9]0.193 & \col[6]0.147 & \col[4]0.101 & {\color{gray}0.088} \\
    $10^{-8}$ & \col[10]0.354 & \col[9]0.237 & \col[9]0.190 & \col[4]0.132 & \col[6]0.103 & {\color{gray}0.090} \\
    $10^{-7}$ & \col[10]0.353 & \col[9]0.228 & \col[9]0.163 & \col[5]0.106 & \col[2]0.094 & {\color{gray}0.100} \\
    $10^{-6}$ & \col[10]0.338 & \col[8]0.202 & \col[5]0.145 & \col[2]0.126 & \col[1]0.122 & {\color{gray}0.150} \\
    $10^{-5}$ & \col[8]0.358 & \col[4]0.295 & \col[1]0.316 & \col[0]0.321 & \col[1]0.326 & {\color{gray}0.434} \\
    $10^{-4}$ & \col[4]0.609 & \col[1]0.833 & \col[0]0.910 & \col[0]0.953 & \col[0]0.935 & {\color{gray}1.155} \\
    $10^{-3}$ & \col[1]2.069 & \col[0]3.202 & \col[0]3.319 & \col[0]3.327 & \col[0]3.341 & {\color{gray}3.454} \\\bottomrule
\end{tabular}
\caption{Average KL divergence (without approximation) from the ground-truth model $\theta_{\text{GT}}$ for MHNs obtained by optimizing the sKL divergence with the static choice of $\teps$ given by \cref{eq:static_eps}.
If $q_{\theta,i}+\eps<0$ occurred during optimization, the last available MHN was used to calculate the KL divergence from the ground truth.
The cell color indicates the percentage of simulations where this occurred (a darker color corresponds to a higher percentage).
In the column ``exact'', the TT format was not used during optimization, thus no negative entries occurred here.}
\label{table:MHNstatic}
\end{table}

For increasing value of $r_\tensor q$, the TT approximation improves, and accordingly the approximate results tend towards the exact result, although the convergence is quite slow.
If $q_{\theta, i}+\eps<0$ occurred in any iteration of the optimization procedure, the optimization was stopped and the last $\theta$ matrix was returned.
The colors indicate the percentage of runs where this happened.

The first row of \cref{table:MHNstatic} shows the results of optimization using the KL divergence.
It can clearly be seen that, when the TT format is used, optimization using the sKL divergence leads to MHNs that describe the data more closely.
The optimal $\eps$ value is between $10^{-7}$ and $10^{-6}$ across all cases where the TT format is used.
The results also clearly show that, as explained in \cref{sec:theory:parameter_choice}, if $\eps$ is chosen too small or too large, we obtain poor optimization results.
From the colors we can see that early stopping due to negative entries of $q_{\theta,i}+\eps$ does not have a big influence on the optimization results.
This is because early stopping often occurred late in the optimization procedure, i.e., after a good estimate for the optimum was already found.

\subsubsection{Dynamic choice} \label{sec:application:dynamic_choice}
\begin{figure}[t]
    \centering
    \includegraphics{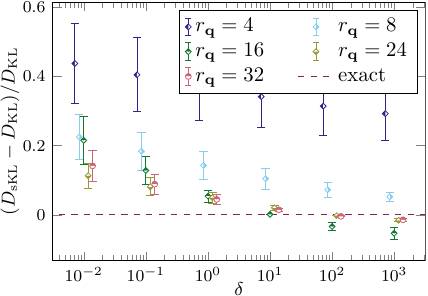}
    \caption{Relative difference between the KL divergence and the sKL divergence with the dynamic choice of $\teps$ given by \cref{eq:dynamic_eps,eq:dynamic_f}, for different TT ranks $r_{\tensor{q}}$.}
    \label{fig:sKLconv}
\end{figure}

Next, we similarly investigate the dynamic choice of $\teps$ given in \cref{eq:dynamic_eps}, choosing the function $f$ as
\begin{equation} \label{eq:dynamic_f}
    f(x)=\delta \cdot x
\end{equation}
with $\delta>0$.
In \cref{thm:noise} we showed that the average of the sKL divergence converges to the KL divergence in the limit of small i.i.d.\ Gaussian noise.
In our particular application, numerical data show that the assumption of i.i.d.\ Gaussian noise is violated.
Nevertheless, as $r_{\tensor q}$ increases and thus the quality of the approximation improves, we observe in \cref{fig:sKLconv} that the sKL divergence converges to the KL divergence, as already mentioned at the end of \cref{sec:theory:dynamic}.
This statement is true for all values of the parameter $\delta$, although the details of the convergence may depend on $\delta$.

\begin{figure}[t]
    \centering
    \includegraphics{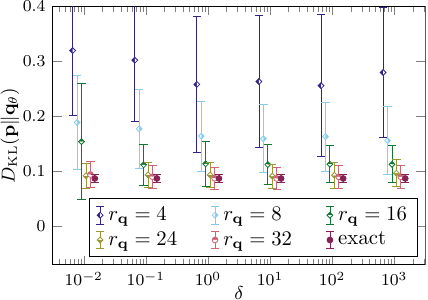}
    \caption{Visualization of the data in \cref{table:MHNdynamic}.}
    \label{fig:MHNdynamic}
\end{figure}

We now discuss the results obtained from optimizing the sKL divergence, similar to \cref{sec:application:static_choice}. Since optimization using higher-order optimizers is not possible with the dynamic choice, simple gradient descent was used for optimization.
The dynamic choice of $\teps$ ensures $q_{\theta, i}+\eps_i>0$, so optimization was always done until a stopping criterion was satisfied.
In \cref{table:MHNdynamic} and \cref{fig:MHNdynamic} we show numerical results for various combinations of $\delta$ and $r_\tensor{q}$.
As $r_\tensor{q}$ is increased, we again observe a convergence towards the exact result, but now at a much faster rate than for the static choice. 
\Cref{table:MHNdynamic} also shows that the results are quite stable with respect to the parameter $\delta$.
For the exact calculation without the TT format, $\delta$ played no role, since it is only important when encountering negative entries in $\tensor{q}_\theta$.

\begin{table}[t]
    \newcommand\col[1][]{\cellcolor{Tpurple!#1}}%
\centering
\tabcolsep=5.5pt
\begin{tabular}{c ccccc c}
    \toprule
    & \multicolumn{5}{c}{$r_\tensor{q}$} & \\\cmidrule{2-6}
    $\delta$ & 4 & 8 & 16 & 24 & 32 & exact\\\midrule
    $10^{-2}$ & \col[40]0.320 & \col[30]0.189 & \col[16]0.154 & \col[2]0.092 & \col[4]0.094 & 0.087 \\
    $10^{-1}$ & \col[60]0.302 & \col[20]0.177 & \col[10]0.112 & \col[2]0.093 & \col[4]0.090 & 0.087 \\
    $1$ & \col[62]0.258 & \col[26]0.164 & \col[26]0.114 & \col[4]0.093 & \col[10]0.087 & 0.087 \\
    $10^{1}$ & \col[74]0.263 & \col[30]0.159 & \col[28]0.112 & \col[10]0.091 & \col[12]0.087 & 0.087 \\
    $10^{2}$ & \col[90]0.256 & \col[28]0.163 & \col[26]0.113 & \col[10]0.093 & \col[12]0.090 & 0.087 \\
    $10^{3}$ & \col[54]0.280 & \col[38]0.156 & \col[28]0.113 & \col[12]0.096 & \col[6]0.090 & 0.087 \\\bottomrule
\end{tabular}
\caption{Average KL divergence (without approximation) from the ground truth model $\theta_{\text{GT}}$ for MHNs obtained by optimizing the sKL divergence with the dynamic choice of $\teps$ given by \cref{eq:dynamic_eps,eq:dynamic_f}.
The colors indicate the number of negative entries $q_{\theta,i}$ in the final results, for indices $i$ with $p_i>0$ (a darker color corresponds to a larger number of negative entries).
In the column ``exact'', the TT format was not used during optimization, thus no negative entries occurred here.}
\label{table:MHNdynamic}
\end{table}

Comparing the static and dynamic choice of $\teps$, it can be seen clearly that dynamically choosing $\teps$ generally leads to MHNs that are closer to the exact results at the same level of approximation.
This is because usually only a few entries of $\tensor{q}_{\theta}$ are negative.
Therefore, dynamically choosing $\teps$ leads to an objective function that closely resembles the KL divergence, while the static choice introduces a shift for all entries even when the shift is not needed.

\section{Summary and Outlook} \label{sec:outlook}
We have introduced a new method to handle negative entries in approximate probability vectors.
This method is very general and can be used in a wide variety of applications (see \cref{sec:related} for examples).
Moreover, it does not come with significant computational overhead. 

We showed that the sKL divergence shares many desirable properties with the KL divergence.
We discussed two possible choices of the shift parameters that occur in the sKL divergence. The static choice allows for the use of higher-order optimizers, but it requires tuning of the parameters and leads to a large loss of gradient information.
In contrast, the dynamic choice restricts us to first-order optimizers, but it offers more freedom in the choice of the parameters and preserves most of the gradient information.
For the dynamic choice we showed that, when negative entries occur due to i.i.d.\ Gaussian noise, the difference between the KL divergence and the average sKL divergence is quadratic in the strength of the noise and thus goes to zero for small noise.

In this work, we only considered the sKL divergence in the context of approximate discrete probability distributions.
The investigation of possible use cases in other contexts is left for future work.

We applied our method to a real-world application, the modeling of cancer progression by Mutual Hazard Networks, where the use of tensor-train approximations can lead to negative entries.
We showed that the sKL divergence and the corresponding model results converge to the KL divergence and the exact model results, respectively, when the parameter that controls the quality of the approximation is increased.
We also showed that the dynamic choice of the shift parameters leads to a faster convergence to the exact results than the static choice, as expected from the theoretical considerations in \cref{sec:theory}. 

When using the sKL divergence as an objective function, first-order optimizers are desirable because they allow for more freedom when choosing the parameters of the sKL divergence.
So far, we only used a standard gradient-descent optimizer.
In future work, we will investigate the effect of different first-order optimizers, including stochastic and momentum-based optimizers.

Regarding MHNs, it was shown in~\cite{Georg_2022_PhD} that the computational complexity of model construction can be reduced from exponential to cubic in the number of events using low-rank tensor formats.
The remaining problem of negative entries in the approximate probability vectors is solved by the current work, which thus enables the construction of large Mutual Hazard Networks with $\gg30$ events.
Expected applications will include as many as $800$ events.
Furthermore, modifications of the MHN have been conceived which allow for more realistic modeling of tumor progression, but which are currently limited to a very small number of events. 
We will apply the techniques described in this work to these large and extended MHNs in future work.

\begin{acknowledgments}
This work is supported in part by the German Research Foundation (DFG) through the grants ``Tensorapproximationsmethoden zur Modellierung von Tumorprogression" (project 458051812), ``PUNCH4NFDI - Teilchen, Universum, Kerne und Hadronen f\"ur die NFDI" (project 460248186), and ``Striking a moving target: From mechanisms of metastatic organ colonization to novel systemic therapies" (TRR 305). We would like to thank Alexander Rothkopf and Phiala Shanahan for a stimulating discussion.
\end{acknowledgments}

\appendix
\allowdisplaybreaks

\section{Proofs} \label{sec:proofs}
\subsection{Proof of \cref{thm:divergence}}
    We first prove property~\hyperlink{thm:div3}{(3)} by direct calculation,
    \begin{align*}
        \frac{\text{d}^2\DsKL(\tensor{p}\|\tensor{q})}{\text{d}q_i\text{d}q_j}\bigg|_{\tensor{q}=\tensor{p}}
        =\frac{\delta_{ij}}{p_i+\eps_i}\,.
    \end{align*}
    This means that ${\text{d}^2\DsKL(\tensor{p}\|\tensor{q})}/{\text{d}q_i\text{d}q_j}|_{\tensor{q}=\tensor{p}}$ are the entries of a diagonal matrix.
    Since all diagonal entries are positive by assumption, the matrix is positive definite.    

    Next, we prove that $\tensor{q}=\tensor{p}$ is the only extremum of $\DsKL$.
    The vectors $\tensor{p}$ and $\tensor{q}$ obey $\sum_ip_i=\sum_iq_i$.
    We can use a Lagrange multiplier $\lambda$ to find the extremal points of
    \begin{align*}
        \mathcal{L}_{\DsKL}(\tensor{q}, \lambda)
        &=\sum_i(p_i+\eps_i)\log\frac{p_i+\eps_i}{q_i+\eps_i}
        + \lambda\sum_i(p_i-q_i)\,.
    \end{align*}
    By taking derivatives w.r.t.\ $q_i$ and $\lambda$, we find that the only local extremum of $\mathcal{L}_{\DsKL}(\tensor{q}, \lambda)$ is at $\tensor{q}=\tensor{p}$ and $\lambda=-1$. 
    Together with property~\hyperlink{thm:div3}{(3)} and $\DsKL(\tensor{p}\|\tensor{p})=0$, this proves properties~\hyperlink{thm:div2}{(2)} and~\hyperlink{thm:div1}{(1)}.

\subsection{Proof of \cref{thm:convexity}}
    With $\lambda_1=\lambda$ and $\lambda_2=1-\lambda$, we can write the left-hand side as
    \begin{align*}
       & \DsKL(\lambda_1 \tensor{p}^{(1)}+\lambda_2 \tensor{p}^{(2)}\|\lambda_1 \tensor{q}^{(1)}+\lambda_2 \tensor{q}^{(2)}) \\
             &=\sum_i\biggl(\sum_{j=1, 2}\lambda_j(p_i^{(j)}+\eps_i)\biggr)
             \log\frac{\sum\limits_{j=1, 2}\lambda_j(p_i^{(j)}+\eps_i)}{\sum\limits_{j=1, 2}\lambda_j(q_i^{(j)}+\eps_i)}\,.
    \end{align*}
    For fixed $i$, we now use the log sum inequality~\cite{Cover_1991}
    \begin{equation*}
        \biggl(\sum_ja_j\biggr)\log\frac{\sum_ja_j}{\sum_jb_j}
        \leq\sum_ja_j\log\frac{a_j}{b_j}
    \end{equation*}
    for $a_j, b_j>0$ to complete the proof.

\subsection{Proof of \cref{thm:noise}} \label{proof:noise}
    The probability distribution of i.i.d.\ Gaussian noise $\tensor{x}$ with mean $0$ and standard deviation $\sigma$ is
    \begin{equation*}
        P(\tensor{x}) = \prod_i\rho(x_i,\sigma) = \prod_i \frac{1}{\sqrt{2\pi\sigma^2}}\exp\left(-\frac{x_i^2}{2\sigma^2}\right).
    \end{equation*}
    The average of the sKL divergence over the Gaussian noise is given by
    \begin{align}
        &\big\langle\DsKL(\tensor{p}\|\tensor{q}+\tensor{x})\big\rangle_\tensor{x} \notag\\
        &= \sum_i\int_{-\infty}^\infty\textup{d} x_i\,\rho(x_i,\sigma)(p_i+\eps_i)\log\frac{p_i+\eps_i}{q_i+x_i+\eps_i}\,. \label{eq:expected_sKL}
    \end{align}
    We therefore consider the integral
    \begin{align*}
        I=c\int_{-\infty}^\infty\textup{d}x\,e^{-\frac{x^2}{2\sigma^2}}(p+\eps)\log\frac{p+\eps}{q+x+\eps}\,,
    \end{align*}
    where $c=1/\sqrt{2\pi\sigma^2}$ is the normalization factor of the Gaussian.
    To expand this integral in $\sigma^2$, it is convenient to perform a saddle-point analysis (also known as method of steepest descent or stationary-phase approximation)~\cite{Mathews_1970}.
    This is a useful method to deal with integrals of the form
    \begin{align*}
        \int \textup{d}x\,e^{-Ng(x)}h(x)\,,
    \end{align*}
    where $N$ is a large parameter and $g$ and $h$ are functions of $x$. The saddle-point method yields a systematic expansion of such integrals in powers of $1/N$.
    In our case, the large parameter is $N=1/\sigma^2$, and $g(x)=x^2/2$. The remaining part of the integrand is identified with $h(x)$.
    
    We first split the integral $I$ into two contributions $I_1$ and $I_2$ corresponding to the integration intervals $(-\infty,-q]$ and $[-q,\infty)$, respectively.
    For $q+x\geq 0$, \cref{eq:dynamic_eps} gives $\eps=0$ so that $I_2$ takes the form
    \begin{align*}
        I_2=c\int_{-q}^{\infty}\textup{d}x\,e^{-\frac{x^2}{2\sigma^2}}\,p\log\frac{p}{q+x} \,.
    \end{align*}
    The saddle point, i.e., the location of the minimum of $g(x)$, is at $x=0$, which is within the integration interval. A standard saddle-point analysis up to next-to-leading order then yields
    \begin{align*}
        I_2=p\log\frac{p}{q}+\frac{p}{2q^2}\sigma^2+\mathcal{O}(\sigma^4)\,.
    \end{align*}
    In the remaining region, where $q+x<0$, \cref{eq:dynamic_eps} gives
    \begin{equation*}
        \eps=|q+x|+f(|q+x|)\,.
    \end{equation*}
    Substituting $x=-q-z$, we  obtain
    \begin{equation*}
    \label{eq:I1}
        I_1=c\int_0^\infty \textup{d}z\,e^{-\frac{(q+z)^2}{2\sigma^2}}(p+z+f(z))\log\frac{p+z+f(z)}{f(z)}\,.
    \end{equation*}
    Before we can expand this integral in $\sigma^2$, we first need to make sure that it exists.
    This depends on the choice of the (nonnegative) function $f$.
    In particular, $I_1$ exists if $f$ is a sum of power laws, as assumed in~\cref{thm:noise}.
    In fact, $I_1$ only diverges for rather exotic choices of $f$ that go to zero too rapidly.
    For example, it diverges at the lower end for $f(z)=\exp(-1/z)$ and at the upper end for $f(z)=\exp(-\exp(z^4))$.
    In the following we restrict ourselves to functions $f$ for which $I_1$ exists. 
    
    To extract the small-$\sigma$ behavior of $I_1$, we again perform a saddle-point analysis. We still have $N=1/\sigma^2$, but now $g(z)=(q+z)^2/2$. 
    In this case, we encounter two complications that require modifications of the saddle-point method.
    The first complication arises because the saddle point obtained from minimizing $g(z)$ is at $z=-q$, which is outside the integration interval.
    Hence the integral is not dominated by the region near a saddle point but by the region near the minimum of $g(z)$ within the integration interval, which is at the lower bound $z=0$, where $g'(0)=q>0$.
    The standard result for this case is obtained by expanding $g(z)$ about zero, which yields
    \begin{align}
    \label{eq:saddle}
        \int_0^\infty \textup{d}z\,e^{-Ng(z)}h(z)=\frac{e^{-Ng(0)}h(0)}{Ng'(0)}\left(1+\mathcal O(1/N)\right)
    \end{align}
    if $h(0)$ is well-defined.
    Applied to our case we find
    \begin{align*}
        I_1=\frac{\sigma}{\sqrt{2\pi}q}e^{-\frac{q^2}{2\sigma^2}}(p+f(0))\log\frac{p+f(0)}{f(0)}\left(1+\mathcal O(\sigma^2)\right),
    \end{align*}
    which is well-defined if $f(0)>0$. By assumption, $f$ is nonnegative, but if $f(0)=0$ a second complication arises since $\log f(0)$ is not defined. In this case, let us assume that the leading behavior near $z=0$ is $f(z)=az^b$ with $a,b>0$.
    If we split the logarithm in the integrand of $I_1$, the integral involving $\log(p+z+f(z))$ is of the form of \cref{eq:saddle} with $h(0)=p\log p$.
    The integral involving $\log f(z)$ is not covered by \cref{eq:saddle}, but it is still dominated by the region near $z=0$.
    For this part, we need the integral (valid for $\mathrm{Re}(\alpha) > -1$)
    \begin{align*}
        &\int_0^\infty \textup{d}z\,e^{-Ng(z)}z^\alpha\log z\\
        &\approx e^{-Ng(0)}\int_0^\infty \textup{d}z\,e^{-Ng'(0)z}z^\alpha\log z\\
        &=\frac{e^{-Ng(0)}\Gamma(\alpha+1)}{[Ng'(0)]^{1+\alpha}}\left(\log\frac1{Ng'(0)}+\Psi(\alpha+1)\right),
    \end{align*}
    where $\Gamma$ and $\Psi$ denote the gamma and digamma function, respectively, and the $\approx$ sign indicates that we have only kept the leading behavior for large $N$ (corresponding to small $\sigma$ below).
    Applying this to our case (with $\alpha=0$, 1, and $b$) and collecting terms, we now find the leading behavior for small $\sigma$,
    \begin{align*}
        I_1\approx \frac{\sigma}{\sqrt{2\pi}q} e^{-\frac{q^2}{2\sigma^2}}p\left(\log\frac pa-b\log\frac{\sigma^2}{q}+b\gamma\right),
    \end{align*}
    where $\gamma=-\Psi(1)$ is Euler's constant. Incidentally, our choice of $f$ in \cref{sec:application:dynamic_choice} corresponds to $a=\delta$ and $b=1$.
    We conclude that for the functions $f$ we admit, the integral $I_1$ is suppressed for small $\sigma$ by a factor of $\exp(-q^2/2\sigma^2)$ and thus goes to zero faster than any power of $\sigma$. (This statement remains true if the leading behavior of $f$ near zero is $f(z)=\exp(-a/z^b)$ with $a>0$ and $b<1$.)
    Adding the two integrals we obtain
    \begin{align*}
        &\big\langle\DsKL(\tensor{p}\|\tensor{q}+\tensor{x})\big\rangle_\tensor{x} \\
        &\quad= \sum_i\left(p_i\log\frac{p_i}{q_i} + \frac{p_i}{2q_i^2}\sigma^2\right)+\mathcal{O}(\sigma^4) \\
        &\quad= \DKL(\tensor{p}\|\tensor{q}) + \sigma^2\sum_i\frac{p_i}{2q_i^2} + \mathcal{O}(\sigma^4)\,,
    \end{align*}
    which completes the proof.

\section{Transition-rate matrix as a tensor train} \label{sec:Q_matrix_TT}
    Here we provide a more detailed description of how the transition-rate matrix is stored in the tensor-train format.
    We start from \cref{eq:MHN_transition_rate_matrix_kron}.
    We can define $4$-dimensional tensors $\tilde Q_{\ell k}\in\mathbb R^{1\times2\times2\times1}$ through $(\tilde Q_{\ell k})_{1ij1}=(Q_{\ell k})_{ij}$
    and construct tensor-train operators $\tilde{\tensor Q}_\ell$ with all TT ranks equal to $1$,
    \begin{align*}
        \tilde{\tensor Q}_\ell(i_1,&\ldots,i_d,j_1,\ldots,j_d)\\
        &=\sum_{\alpha_0,\ldots,\alpha_d}\prod_{k=1}^d\tilde Q_{\ell k}(\alpha_{k-1},i_k,j_k,\alpha_k)\,.
    \end{align*}
    The full transition-rate matrix is now obtained in the TT format by summing up the tensor trains $\tilde{\tensor Q}_\ell$, giving
    \begin{equation*}
        \tilde{\tensor Q}(i_1,\ldots,i_d,j_1,\ldots,j_d)=\sum_{\ell=1}^d \tilde{\tensor Q}_\ell(i_1,\ldots,i_d,j_1,\ldots,j_d)\,.
    \end{equation*}
    This sum can be performed in the format without approximation, resulting in a tensor train with all TT ranks equal to $d$ (except for $r_0=r_d=1$)~\cite{Hackbusch_2019,Oseledets_2011}.

\bibliography{bib}

\end{document}